\documentstyle[12pt]{article}
\newlength{\extraspace}
\newlength{\extraspaces}
\newcommand{\be}{\begin{equation}
\addtolength{\abovedisplayskip}{\extraspaces}
\addtolength{\belowdisplayskip}{\extraspaces}
\addtolength{\abovedisplayshortskip}{\extraspace}
\addtolength{\belowdisplayshortskip}{\extraspace}}
\newcommand{\ee}{\end{equation}}

\newcommand{\ba}{\begin{eqnarray}
\addtolength{\abovedisplayskip}{\extraspaces}
\addtolength{\belowdisplayskip}{\extraspaces}
\addtolength{\abovedisplayshortskip}{\extraspace}
\addtolength{\belowdisplayshortskip}{\extraspace}}
\newcommand{\ea}{\end{eqnarray}}

\newcommand{\newsection}[1]{
\vspace{15mm}
\pagebreak[3]
\addtocounter{section}{1}
\setcounter{equation}{0}
\setcounter{subsection}{0}
\setcounter{footnote}{0}
\begin{flushleft}
{\large\bf \thesection. #1}
\end{flushleft}
\nopagebreak
\medskip
\nopagebreak}

\newcommand{\Tr}{{\rm Tr}}

\begin{document}

\addtolength{\baselineskip}{.8mm}

{\thispagestyle{empty}
\begin{center}\vspace*{1.0cm}
{\large\bf The analytic continuation of the high--energy} \\
{\large\bf quark--quark scattering amplitude} \\
\vspace*{1.0cm}
{\large Enrico Meggiolaro} \\
\vspace*{0.5cm}{\normalsize
{Dipartimento di Fisica, \\
Universit\`a di Pisa, \\
and INFN, Sezione di Pisa, \\ 
I--56100 Pisa, Italy.}} \\
\vspace*{2cm}{\large \bf Abstract}
\end{center}

\noindent
It is known that the high--energy quark--quark scattering amplitude can be
described by the expectation value of two lightlike Wilson lines, running 
along the classical trajectories of the two colliding particles.
Generalizing the results of a previous paper, we give here the general proof 
that the expectation value of two infinite Wilson lines, forming a certain 
hyperbolic angle in Minkowski space--time, and the expectation value of two 
infinite Euclidean Wilson lines, forming a certain angle in Euclidean 
four--space, are connected by an analytic continuation in the angular 
variables. This result could be used to evaluate the high--energy scattering 
amplitude directly on the lattice.
}
\vfill\eject

\newsection{Introduction}

\noindent
It is well known that the quark--quark scattering amplitude, at high squared 
energies $s$ in the center of mass and small squared transferred momentum $t$ 
(that is $s \to \infty$ and $|t| \ll s$, let us say $|t| \le 1~{\rm GeV}^2$), 
can be described by the expectation value of two lightlike Wilson lines, 
running along the classical trajectories of the two colliding particles
\cite{Nachtmann91} \cite{Meggiolaro96}.

In the center--of--mass reference system (c.m.s.), taking for example the 
initial trajectories of the two quarks along the $x^1$--axis, 
the scattering amplitude has the following form 
[explicitly indicating the color indices ($i,j, \ldots$)
and the spin indices ($\alpha, \beta, \ldots$) of the quarks]
\ba
\lefteqn{
M_{fi} = \langle \psi_{i\alpha}(p'_1) \psi_{k\gamma}(p'_2) | M | 
\psi_{j\beta}(p_1) \psi_{l\delta}(p_2) \rangle } \nonumber \\
& & \mathop{\sim}_{s \to \infty}
-{i \over Z_\psi^2} \cdot \delta_{\alpha\beta} \delta_{\gamma\delta}
\cdot 2s 
\displaystyle\int d^2 {\bf z}_t e^{i {\bf q} \cdot {\bf z}_t}
\langle [ W_1 (z_t) - {\bf 1} ]_{ij} [ W_2 (0) - {\bf 1} ]_{kl} \rangle
~,
\ea
where $q = (0,0,{\bf q})$, with $t = q^2 = -{\bf q}^2$, is the tranferred 
four--momentum and $z_t = (0,0,{\bf z}_t)$, with ${\bf z}_t = (z^2,z^3)$, 
is the distance between the two trajectories in the {\it transverse} plane
[the coordinates $(x^0,x^1)$ are often called {\it longitudinal} coordinates].
The expectation 
value $\langle f(A) \rangle$ is the average of $f(A)$ in the sense of the 
functional integration over the gluon field $A^\mu$ (including also the 
determinant of the fermion matrix, i.e., $\det[i\gamma^\mu D_\mu - m]$,
where $D^\mu = \partial^\mu + ig A^\mu$ is the covariant derivative)
\cite{Nachtmann91} \cite{Meggiolaro96}.
The two lightlike Wilson lines $W_1 (z_t)$ and $W_2 (0)$ in Eq. (1.1) are 
defined as
\ba
W_1 (z_t) &=&
{P} \exp \left[ -ig \displaystyle\int_{-\infty}^{+\infty}
A_\mu (z_t + p_1 \tau) p_1^\mu d\tau \right] ~;
\nonumber \\
W_2 (0) &=&
{P} \exp \left[ -ig \displaystyle\int_{-\infty}^{+\infty}
A_\mu (p_2 \tau) p_2^\mu d\tau \right] ~,
\ea
where $P$ stands for ``{\it path ordering}'' and $A_\mu = A_\mu^a T^a$;
$p_1 \simeq (E,E,{\bf 0}_t)$ and $p_2 \simeq (E,-E,{\bf 0}_t)$ are the 
initial four--momenta of the two quarks.
The space--time configuration of these two Wilson lines is shown in Fig. 1.

\newcommand{\wilson}[1]{
\begin{figure}
\begin{center}
\setlength{\unitlength}{1.00mm}
\raisebox{-40\unitlength}
{\mbox{\begin{picture}(80,45)(-35,-30)
\thicklines
\put(-22,22){\line(1,-1){41}}
\put(-15,15){\vector(-1,1){1}}
\put(16,23){\line(-1,-1){42}}
\put(9,16){\vector(1,1){1}}
\put(0,0){\vector(-2,-1){13}}
\thinlines
\put(-8,0){\line(1,0){35}}
\put(8,4){\line(-2,-1){35}}
\put(0,-8){\line(0,1){35}}
\put(27,0){\vector(1,0){1}}
\put(0,27){\vector(0,1){1}}
\put(-13,20){\makebox(0,0){$W_2$}}
\put(18,20){\makebox(0,0){$W_1$}}
\put(-6,-7){\makebox(0,0){$z_t$}}
\put(25,-2){\makebox(0,0){$x$}}
\put(-2,25){\makebox(0,0){$t$}}
\end{picture}}}
\parbox{13cm}{\small #1}
\end{center}
\end{figure}}

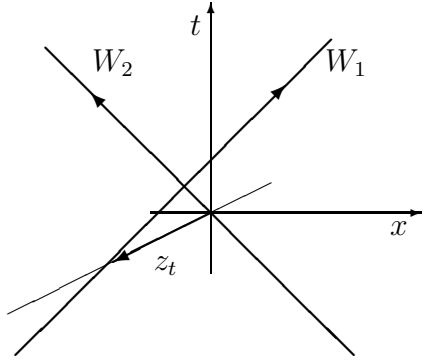
\wilson{{\bf Fig.~1.} The space--time configuration of the two lightlike
Wilson lines $W_1$ and $W_2$ entering in the expression (1.1) for the
high--energy quark--quark elastic scattering amplitude.}

Finally, $Z_\psi$ in Eq. (1.1) is the fermion--field renormalization constant, 
which can be written in the eikonal approximation as \cite{Nachtmann91} 
\be
Z_\psi \simeq {1 \over N_c} \langle \Tr [ W_1 (z_t) ] \rangle
= {1 \over N_c} \langle \Tr [ W_1 (0) ] \rangle
= {1 \over N_c} \langle \Tr [ W_2 (0) ] \rangle ~,
\ee
where $N_c$ is the number of colours.

In what follows, we shall deal with the quantity
\be
g_{M (ij,kl)} (s; ~t) \equiv {1 \over Z_\psi^2}
\displaystyle\int d^2 {\bf z}_t e^{i {\bf q} \cdot {\bf z}_t}
\langle [ W_1 (z_t) - {\bf 1} ]_{ij} [ W_2 (0) - {\bf 1} ]_{kl} \rangle ~,
\ee
in terms of which the scattering amplitude can be written as
\be
M_{fi} = \langle \psi_{i\alpha}(p'_1) \psi_{k\gamma}(p'_2) | M | 
\psi_{j\beta}(p_1) \psi_{l\delta}(p_2) \rangle
\mathop{\sim}_{s \to \infty}
-i \cdot 2s \cdot \delta_{\alpha\beta} \delta_{\gamma\delta}
\cdot g_{M (ij,kl)} (s; ~t) ~.
\ee
The quantity $g_{M (ij,kl)} (s; ~t)$ depends not only on $t = -{\bf q}^2$,
but also on $s$. In fact, as was pointed out by Verlinde and Verlinde in 
\cite{Verlinde}, it is a singular limit to take the Wilson lines in
(1.4) exactly lightlike. A way to regularize this sort of ``infrared'' 
divergence (so called because it essentially comes from the limit $m \to 0$,
where $m$ is the quark mass) consists in letting each line
have a small timelike component, so that they coincide with the classical 
trajectories for quarks with a finite mass $m$ (see also Ref. \cite{zfp97} 
for a discussion about this point). In other words, one first evaluates
the quantity $g_{M (ij,kl)} (\beta; ~t)$ for two Wilson lines along the 
trajectories of two quarks (with mass $m$) moving with velocity $\beta$ and 
$-\beta$ ($0 <  \beta <  1$) in the $x^1$--direction.
This is equivalent to consider two infinite Wilson lines forming a 
certain (finite) hyperbolic angle $\chi$ in Minkowski space--time.
Then, to obtain the correct high--energy scattering amplitude, one has to 
perform the limit $\beta \to 1$, that is $\chi \to \infty$, in the 
expression for $g_{M (ij,kl)} (\beta; ~t)$:
\be
M_{fi} = \langle \psi_{i\alpha}(p'_1) \psi_{k\gamma}(p'_2) | M | 
\psi_{j\beta}(p_1) \psi_{l\delta}(p_2) \rangle
\mathop{\sim}_{s \to \infty}
-i \cdot 2s \cdot \delta_{\alpha\beta} \delta_{\gamma\delta}
\cdot g_{M (ij,kl)} (\beta \to 1; ~t) ~.
\ee
Proceeding in this way one obtains 
a $\ln s$ dependence of the amplitude, as expected from ordinary 
perturbation theory and as confirmed by the experiments on hadron--hadron 
scattering processes \cite{Cheng-Wu-book} \cite{Lipatov}.
In Sect. 3 of Ref. \cite{zfp97} we have followed this procedure to explicitly 
evaluate the second member of (1.6) up to the fourth order in the expansion 
in the renormalized coupling constant: the results so derived are in 
agreement with those obtained from ordinary perturbation theory.

The direct evaluation of the expectation value (1.4) is a 
highly non--trivial matter and it is strictly connected with the
ultraviolet properties of Wilson--line operators 
\cite{Arefeva80} \cite{Korchemsky}.
Some non--perturbative approaches for the calculation of (1.4) have been 
proposed in Refs. \cite{Arefeva94} and \cite{Dosch}.

In a recent paper \cite{zfp97} we have proposed a new approach, which 
consists in adapting the scattering amplitude to the Euclidean world:
this approach could open the way for the direct evaluation of the scattering 
amplitude on the lattice.
More explicitly, in Ref. \cite{zfp97} we have given arguments showing 
that the expectation value of two infinite Wilson lines,
forming a certain hyperbolic angle in Minkowski space--time, and the 
expectation value of two infinite Euclidean Wilson lines, forming a certain 
angle in Euclidean four--space, are likely to be connected by an analytic
continuation in the angular variables. 
This relation of analytic continuation has been proven in Ref. \cite{zfp97} 
for an Abelian gauge theory (QED) in the so--called {\it quenched} 
approximation and for a non--Abelian gauge theory (QCD) up to the fourth 
order in the renormalized coupling constant in perturbation theory:
a general proof was missing up to now.

In this paper, we shall generalize the results of Ref. \cite{zfp97}
and give the rigorous proof of the above--mentioned relation of 
analytic continuation for a non--Abelian gauge theory with gauge group 
$SU(N_c)$ [as well as for an Abelian gauge theory (QED)]. 
The approach adopted in Ref. \cite{zfp97} consisted in explicitly
evaluating the amplitudes $g_M (\chi; ~t)$ and $g_E (\theta; ~t)$, in the 
Minkowski and the Euclidean world, in some given approximation (such as the
{\it quenched} approximation) or up to some order in perturbation theory and 
in finally comparing the two expressions so obtained. Instead, in this 
paper we shall give a general proof, which essentially exploits the relation 
between the gluonic Green functions in the two theories.
\vfill\eject

\newsection{From Minkowskian to Euclidean theory}

\noindent
Let us consider the following quantity, defined in
Minkowski space--time:
\ba
g_M (p_1, p_2; ~t) &=& {M (p_1, p_2; ~t) \over Z_W^2} ~, \nonumber \\
M (p_1, p_2; ~t) &=& 
\displaystyle\int d^2 {\bf z}_t e^{i {\bf q} \cdot {\bf z}_t}
\langle [ W_1 (z_t) - {\bf 1} ]_{ij} [ W_2 (0) - {\bf 1} ]_{kl} \rangle ~,
\ea
where $p_1$ and $p_2$ are the four--momenta [lying (for example) in the plane 
$(x^0,x^1)$], which define the trajectories of the two Wilson lines $W_1$ and 
$W_2$ ($A_\mu = A_\mu^a T^a$ and $m$ is the fermion mass):
\ba
W_1 (z_t) &\equiv&
{P} \exp \left[ -ig \displaystyle\int_{-\infty}^{+\infty}
A_\mu (z_t + {p_1 \over m} \tau) {p_1^\mu \over m} d\tau \right] ~;
\nonumber \\
W_2 (0) &\equiv&
{P} \exp \left[ -ig \displaystyle\int_{-\infty}^{+\infty}
A_\mu ({p_2 \over m} \tau) {p_2^\mu \over m} d\tau \right] ~.
\ea
$Z_W$ in Eq. (2.1) is defined as ($N_c$ being the number of colours)
\be
Z_W \equiv {1 \over N_c} \langle \Tr [ W_1 (z_t) ] \rangle
= {1 \over N_c} \langle \Tr [ W_1 (0) ] \rangle = {1 \over N_c}
\langle \Tr [ W_2 (0) ] \rangle ~.
\ee
(The two last equalities come from the Poincar\'e invariance.)
This is a sort of Wilson--line's renormalization constant:
as shown in Ref. \cite{Nachtmann91}, $Z_W$ coincides with the fermion 
renormalization constant $Z_\psi$ in the eikonal approximation.

By virtue of the Lorentz symmetry, we can define $p_1$ and $p_2$ in the 
c.m.s. of the two particles, moving with speed $\beta$ and $-\beta$ along 
the $x^1$--direction:
\ba
p_1^\mu &=& E (1,\beta,{\bf 0}_t) ~, \nonumber \\
p_2^\mu &=& E (1,-\beta,{\bf 0}_t) ~, 
\ea
where $E = m / \sqrt{1 - \beta^2}$ (in units with $c=1$) is the 
energy of each particle (so that: $s = 4E^2$).

We now introduce the hyperbolic angle $\psi$ [in the plane $(x^0,x^1)$]
of the trajectory of $W_1$: it is given by $\beta = \tanh \psi$.
We can give the explicit form of the Minkowski four--vectors
$u_1 = p_1/m$ and $u_2 = p_2/m$ in terms of the hyperbolic angle $\psi$:
\ba
u_1 = {p_1 \over m} &=& (\cosh \psi,\sinh \psi,{\bf 0}_t) ~, \nonumber \\
u_2 = {p_2 \over m} &=& (\cosh \psi,-\sinh \psi,{\bf 0}_t) ~.
\ea
Clearly, $u_1^2 = u_2^2 = 1$ and
\be
u_1 \cdot u_2 = \cosh (2\psi) = \cosh \chi ~,
\ee
where $\chi = 2\psi$ is the hyperbolic angle [in the plane $(x^0,x^1)$] 
between the two trajectories of $W_1$ and $W_2$.

In an analogous way, we can consider the following quantity, defined
in Euclidean space--time:
\ba
g_E (p_{1E}, p_{2E}; ~t) &=& {E (p_{1E}, p_{2E}; ~t) \over Z_{W E}^2} ~,
\nonumber \\
E (p_{1E}, p_{2E}; ~t) &=& 
\displaystyle\int d^2 {\bf z}_t e^{i {\bf q} \cdot {\bf z}_t}
\langle [ W_{1 E} (z_{t E}) - {\bf 1} ]_{ij} [ W_{2 E} (0) - {\bf 1} ]_{kl} 
\rangle_E ~,
\ea
where $z_{t E} = (z_1, z_2, z_3, z_4) = (0, {\bf z}_t, 0)$ and
$q_E = (0, {\bf q}, 0)$ (so that: $q_E^2 = {\bf q}^2 = -t$).
The expectation value $\langle \ldots \rangle_E$ must be intended now as a 
functional integration with respect to the gauge variable $A^{(E)}_\mu = 
A^{(E)a}_\mu T^a$ in the Euclidean theory.
The Euclidean four--vectors $p_{1E}$ and $p_{2E}$ [lying (for example) in the 
plane $(x_1,x_4)$] define the trajectories of the two Euclidean Wilson lines 
$W_{1 E}$ and $W_{2 E}$: 
\ba
W_{1 E} (z_{t E}) &\equiv&
{P} \exp \left[ -ig \displaystyle\int_{-\infty}^{+\infty}
A^{(E)}_{ \mu} (z_{t E} + p_{1E} \tau) p_{1E \mu} d\tau \right] ~;
\nonumber \\
W_{2 E} (0) &\equiv&
{P} \exp \left[ -ig \displaystyle\int_{-\infty}^{+\infty}
A^{(E)}_{ \mu} (p_{2E} \tau) p_{2E \mu} d\tau \right] ~.
\ea
$Z_{W E}$ in Eq. (2.7) is defined analogously to $Z_W$ in Eq. (2.3):
\be
Z_{W E} \equiv {1 \over N_c} \langle \Tr [ W_{1 E} (z_{t E}) ] \rangle
= {1 \over N_c} \langle \Tr [ W_{1 E} (0) ] \rangle = {1 \over N_c}
\langle \Tr [ W_{2 E} (0) ] \rangle ~.
\ee
(The two last equalities come from the $O(4)$ {\it plus} 
translation invariance.)

We can now expand the Wilson lines $W_1$ and $W_2$ in power series of the 
coupling constant $g$ and take the pieces with $g^n$ and $g^r$ respectively.
Their contribution to the amplitude $M(p_1, p_2; ~t)$ will be called
$M_{(n,r)} (p_1, p_2; ~t)$ (so that $M = \sum_{n=1}^\infty \sum_{r=1}^\infty 
M_{(n,r)}$) and is given by
\ba
\lefteqn{
M_{(n,r)} (p_1, p_2; ~t) = 
(-ig)^{(n+r)} (T^{a_1} \ldots T^{a_n})_{ij} 
(T^{b_1} \ldots T^{b_r})_{kl}
\displaystyle\int d^2 {\bf z}_t e^{i {\bf q} \cdot {\bf z}_t} \times
} \nonumber \\
& & \displaystyle\int d\tau_1 {p_1^{\mu_1} \over m} \ldots 
\displaystyle\int d\tau_n {p_1^{\mu_n} \over m} 
\displaystyle\int d\omega_1 {p_2^{\nu_1} \over m} \ldots
\displaystyle\int d\omega_r {p_2^{\nu_r} \over m} \times
\nonumber \\
& & \theta (\tau_n - \tau_{n-1}) \ldots \theta (\tau_2 - \tau_1)
\theta (\omega_r - \omega_{r-1}) \ldots \theta (\omega_2 - \omega_1) \times
\nonumber \\
& & \langle A_{\mu_1}^{a_1} (z_t + {p_1 \over m} \tau_1) \ldots
A_{\mu_n}^{a_n} (z_t + {p_1 \over m} \tau_n)
A_{\nu_1}^{b_1} ({p_2 \over m} \omega_1) \ldots 
A_{\nu_r}^{b_r} ({p_2 \over m} \omega_r) \rangle ~.
\ea
The corresponding quantity for the Euclidean theory $E_{(n,r)} (p_{1E}, 
p_{2E}; ~t)$, obtained taking the pieces with $g^n$ and $g^r$ in the 
expansion of the Euclidean Wilson lines $W_{1 E}$ and $W_{2 E}$ inside
$E(p_{1E}, p_{2E}; ~t)$, is given by
\ba
\lefteqn{
E_{(n,r)} (p_{1E}, p_{2E}; ~t) =
(-ig)^{(n+r)} (T^{a_1} \ldots T^{a_n})_{ij} 
(T^{b_1} \ldots T^{b_r})_{kl}
\displaystyle\int d^2 {\bf z}_t e^{i {\bf q} \cdot {\bf z}_t} \times
} \nonumber \\
& & \displaystyle\int d\tau_1 p_{1E}^{\mu_1} \ldots 
\displaystyle\int d\tau_n p_{1E}^{\mu_n}
\displaystyle\int d\omega_1 p_{2E}^{\nu_1} \ldots
\displaystyle\int d\omega_r p_{2E}^{\nu_r} 
\theta (\tau_n - \tau_{n-1}) \ldots \theta (\tau_2 - \tau_1) \times
\nonumber \\
& & \theta (\omega_r - \omega_{r-1}) \ldots \theta (\omega_2 - \omega_1)
\langle A_{(E) \mu_1}^{a_1} (z_{t E} + p_{1E} \tau_1) \ldots
A_{(E) \mu_n}^{a_n} (z_{t E} + p_{1E} \tau_n) \times
\nonumber \\
& & \times A_{(E) \nu_1}^{b_1} (p_{2 E} \omega_1) \ldots 
A_{(E) \nu_r}^{b_r} (p_{2 E} \omega_r) \rangle_E ~.
\ea
It is known that, making use of the correspondence
\ba
A_0 (x) \rightarrow i A^{(E)}_{ 4}  (x_E) ~~ &,& ~~
A_k (x) \rightarrow A^{(E)}_{ k} (x_E) ~~~ \nonumber \\
{\rm with:}~~x^0 \rightarrow -i x_{E 4} ~~ &,& ~~ 
{\bf x} \rightarrow {\bf x}_E ~,
\ea
between the Minkowski and the Euclidean world, the following relationship 
is derived between the gluonic Green functions in the two theories:
\ba
\lefteqn{
\tilde{B}_{(1)}^{\mu_1} \ldots \tilde{B}_{(N)}^{\mu_N} 
\langle A_{\mu_1}^{a_1} (\tilde{x}_{(1)}) \ldots
A_{\mu_N}^{a_N} (\tilde{x}_{(N)}) \rangle = } \nonumber \\
& & = B_{(1) E \mu_1} \ldots B_{(N) E \mu_N} 
\langle A_{(E) \mu_1}^{a_1} (x_{(1) E}) \ldots
A_{(E) \mu_N}^{a_N} (x_{(N) E}) \rangle_E ~,
\ea
where $x_{(i) E} = ({\bf x}_{(i) E}, x_{(i) E 4})$ are Euclidean 
four--coordinates and $B_{(i) E} = ({\bf B}_{(i) E}, B_{(i) E 4})$ are any 
Euclidean four--vectors, while $\tilde{x}_{(i)}$ and $\tilde{B}_{(i)}$ are 
Minkowski four--vectors defined as
\ba
\tilde{x}_{(i)} = (\tilde{x}_{(i)}^0, \tilde{\bf x}_{(i)}) &=&
(-i x_{(i) E 4}, {\bf x}_{(i) E}) ~, \nonumber \\
\tilde{B}_{(i)} = (\tilde{B}_{(i)}^0, \tilde{\bf B}_{(i)}) &=&
(-i B_{(i) E 4}, {\bf B}_{(i) E}) ~.
\ea
For example, in the case $N = 2$, if one defines the gluonic propagators as
\ba
G_{\mu\nu}^{ab} (x,y) &\equiv& \langle A_\mu^a (x) A_\nu^b (y) \rangle ~,
\nonumber \\
G_{(E) \mu\nu}^{ab} (x_E,y_E) &\equiv& \langle A_{(E) \mu}^a (x_E) 
A_{(E) \nu}^b (y_E) \rangle_E ~,
\ea
one finds that
\ba
G_{00}^{ab} (\tilde{x},\tilde{y}) &=& -G_{(E) 44}^{ab} (x_E,y_E) ~,
\nonumber \\
G_{0j}^{ab} (\tilde{x},\tilde{y}) &=& i G_{(E) 4j}^{ab} (x_E,y_E) ~,
\nonumber \\
G_{j0}^{ab} (\tilde{x},\tilde{y}) &=& i G_{(E) j4}^{ab} (x_E,y_E) ~,
\nonumber \\
G_{jk}^{ab} (\tilde{x},\tilde{y}) &=& G_{(E) jk}^{ab} (x_E,y_E) ~,
\ea
where $j,k = 1,2,3$ are indices for the spatial components and $\tilde{x}$ 
and $\tilde{y}$ are defined as in Eq. (2.14). From these relations, one 
immediately derives Eq. (2.13) for $N = 2$, with $\tilde{B}$ defined as in
Eq. (2.14). The result can be trivially generalized to every $N$.

In our specific case, we can use Eq. (2.13) to state that
\ba
\lefteqn{
{\tilde{p}_1^{\mu_1} \over m} \ldots {\tilde{p}_1^{\mu_n} \over m} 
{\tilde{p}_2^{\nu_1} \over m} \ldots {\tilde{p}_2^{\nu_r} \over m}
\langle A_{\mu_1}^{a_1} (z_t + {\tilde{p}_1 \over m} \tau_1) 
\ldots A_{\mu_n}^{a_n} (z_t + {\tilde{p}_1 \over m} \tau_n)
A_{\nu_1}^{b_1} ({\tilde{p}_2 \over m} \omega_1) \ldots 
A_{\nu_r}^{b_r} ({\tilde{p}_2 \over m} \omega_r) \rangle = }
\nonumber \\
& & = {p_{1E}^{\mu_1} \over m} \ldots {p_{1E}^{\mu_n} \over m}
{p_{2E}^{\nu_1} \over m} \ldots {p_{2E}^{\nu_r} \over m} 
\langle A_{(E) \mu_1}^{a_1} (z_{t E} + {p_{1E} \over m} \tau_1) 
\ldots A_{(E) \mu_n}^{a_n} (z_{t E} + {p_{1E} \over m} \tau_n) \times
\nonumber \\
& & \times A_{(E) \nu_1}^{b_1} ({p_{2 E} \over m} \omega_1) \ldots 
A_{(E) \nu_r}^{b_r} ({p_{2 E} \over m} \omega_r) \rangle_E ~,
\ea
where $p_{iE} = ({\bf p}_{iE}, p_{iE4})$, for $i = 1,2$, are two Euclidean
four--vectors and $\tilde{p}_i$ are the two corresponding Minkowski
four--vectors, obtained according to Eq. (2.14):
\be
\tilde{p}_i = (\tilde{p}_i^0, {\tilde{\bf p}}_i) =
(-i p_{iE4}, {\bf p}_{iE}) ~.
\ee
By virtue of the definitions (2.10) and (2.11) for $M_{(n,r)}$ and $E_{(n,r)}$ 
respectively, Eq. (2.17) implies that:
\be
E_{(n,r)} ({p_{1E} \over m}, {p_{2E} \over m}; ~t) =
M_{(n,r)} (\tilde{p}_1, \tilde{p}_2; ~t) ~.
\ee
This relation is valid for every couple of integer numbers $(n,r)$, so that, 
more generally:
\be
E ({p_{1E} \over m}, {p_{2E} \over m}; ~t) =
M (\tilde{p}_1, \tilde{p}_2; ~t) ~.
\ee
Of course $M$, considered as a general function of $p_1$, $p_2$ [and
$q = (0,0,{\bf q})$], can only depend on the scalar quantities constructed 
with the vectors $p_1$, $p_2$ and $q = (0,0,{\bf q})$: the only 
possibilities are $q^2 = -{\bf q}^2 = t$, $p_1 \cdot p_2$, $p_1^2$ and 
$p_2^2$, since $p_1 \cdot q = p_2 \cdot q = 0$.
Moreover, it is clear from the definitions (2.1) and (2.2) that $M$ cannot 
depend on the (positive) normalizations of the four--vectors $p_1$ and $p_2$: 
in other words, we obtain the same result for $M$ if we substitute 
$(p_1, p_2)$ with $(\alpha_1 p_1, \alpha_2 p_2)$, $\alpha_1$ and $\alpha_2$ 
being arbitrary positive constants.

Therefore, $M$ is forced to have the following form: 
\be
M (p_1, p_2; ~t) = f_M \left( {p_1 \over \sqrt{p_1^2}} \cdot 
{p_2 \over \sqrt{p_2^2}}; ~t \right) ~.
\ee
For analogous reasons, $E$ must be of the form:
\be
E (p_{1E}, p_{2E}; ~t) = f_E \left( {p_{1E} \over |p_{1E}|} \cdot 
{p_{2E} \over |p_{2E}|}; ~t \right) ~,
\ee
where $|B_E| \equiv \sqrt{\sum_{\mu = 1}^4 B_{E\mu}^2}$ is the Euclidean 
norm. (A short remark about the notation: we have denoted everywhere the 
scalar product by a ``$\cdot$'', both in the Minkowski and the Euclidean 
world. Of course, when $A$ and $B$ are Minkowski four--vectors, then
$A \cdot B = A^\mu B_\mu = A^0 B^0 - {\bf A} \cdot {\bf B}$; while, if
$A_E$ and $B_E$ are Euclidean four--vectors, then
$A_E \cdot B_E = A_{E \mu} B_{E \mu} = {\bf A}_E \cdot {\bf B}_E + 
A_{E 4} B_{E 4}$.)
Therefore, the relation (2.20) can be re--formulated as follows [observing 
that $(p_{iE}/m) / |(p_{iE}/m)| = p_{iE} / |p_{iE}|$]
\be
f_E (v_{1E} \cdot v_{2E}; ~t) = f_M (\bar{u}_1 \cdot \bar{u}_2; ~t) ~,
\ee
where $v_{1E}$ and $v_{2E}$ are the Euclidean four--versors corresponding 
to $p_{1E}$ and $p_{2E}$ ($v_{1E}^2 = v_{2E}^2 = 1$):
\be
v_{1E} = {p_{1E} \over |p_{1E}|} ~~~ , ~~~
v_{2E} = {p_{2E} \over |p_{2E}|} ~,
\ee
while $\bar{u}_1$ and $\bar{u}_2$ are the Minkowski four--vectors
defined as
\be
\bar{u}_1 = {\tilde{p}_1 \over \sqrt{\tilde{p}_1^2}} ~~~ , ~~~
\bar{u}_2 = {\tilde{p}_2 \over \sqrt{\tilde{p}_2^2}} ~.
\ee
(It is clear that: $\bar{u}_1^2 = \bar{u}_2^2 = 1$.)
By virtue of the $O(4)$ symmetry of the Euclidean theory, we can choose
a reference frame in which the spatial vectors ${\bf v}_{1E}$ and 
${\bf v}_{2E} = -{\bf v}_{1E}$ are along the $x_1$--direction. 
The two four--momenta $v_{1E}$ and $v_{2E}$ are, therefore,
\ba
v_{1E} &=& (\sin \phi, {\bf 0}_t, \cos \phi ) ~; \nonumber \\
v_{2E} &=& (-\sin \phi, {\bf 0}_t, \cos \phi ) ~,
\ea
where $\phi$ is the angle formed by each trajectory with the $x_4$--axis.
The value of $\phi$ is between $0$ and $\pi / 2$, so that the angle
$\theta = 2 \phi$ between the two Euclidean trajectories $W_{1E}$ and 
$W_{2E}$ lies in the range $[0,\pi]$: it is always possible to make such 
a choice by virtue of the $O(4)$ symmetry of the Euclidean theory.
In such a reference frame, we can write $v_{1E} \cdot v_{2E} = \cos \theta$.

From Eq. (2.18) we have that $\tilde{p}_i^2 = -|p_{iE}|^2 < 0$ and
$\sqrt{\tilde{p}_i^2} = -i |p_{iE}|$. The sign of the squared root is fixed 
in the following way: in the system where ${\bf p}_i = {\bf 0}$, we have 
that $\sqrt{p_i^2} = p_i^0$ (if we take $p_i^0 > 0$).
This relation is continued so to have 
$\sqrt{\tilde{p}_i^2} = \tilde{p}_i^0$ in the system where 
$\tilde{\bf p}_i = {\bf 0}$. But $\tilde{\bf p}_i = {\bf p}_{iE} = {\bf 0}$,
so that $\tilde{p}_i^0 = -i p_{iE4} = -i |p_{iE}|$  
(if we take $p_{iE4} > 0$). Therefore, in this particular system
$\sqrt{\tilde{p}^2} = \tilde{p}^0 = -i p_{iE4} = -i |p_{iE}|$.
So we take
$\sqrt{\tilde{p}^2} = -i |p_{iE}|$ 
in every system. This implies that:
\be
\bar{u}_i = {\tilde{p}_i \over \sqrt{\tilde{p}_i^2}} =
(v_{iE4}, i{\bf v}_{iE}) ~.
\ee
With the explicit form of $v_{1E}$ and $v_{2E}$ given by Eq. (2.26), we find 
that
\ba
\bar{u}_1 &=& (\cos \phi,i\sin \phi,{\bf 0}_t) ~, \nonumber \\
\bar{u}_2 &=& (\cos \phi,-i\sin \phi,{\bf 0}_t) ~,
\ea
and consequently $\bar{u}_1^2 = \bar{u}_2^2 = 1$ and
\be
\bar{u}_1 \cdot \bar{u}_2 = \cos (2\phi) = \cos \theta ~.
\ee
A comparison with the expressions (2.5) for the Minkowski four--vectors
$u_1$ and $u_2$ reveals that $\bar{u}_1$ and $\bar{u}_2$ are obtained from
$u_1$ and $u_2$ after the following analytic continuation in the angular 
variables is made:
\be
\chi \rightarrow i \theta ~.
\ee
(We remind that $\phi = \theta/2$ and $\psi = \chi/2$.)
Therefore, by virtue of Eqs. (2.21) and (2.22), the relation (2.23) can be 
formulated as follows:
\be
E (\theta; ~t) = M (\chi \to i\theta; ~t) ~.
\ee
Let us consider, now, the Wilson--line's renormalization constant $Z_W$:
\be
Z_W \equiv {1 \over N_c} \langle \Tr [ W_1 (0) ] \rangle ~.
\ee
We can expand $W_1$ in power series of $g$ and take the piece with $g^n$,
whose contribution to $Z_W$ we call $Z_W^{(n)}$:
\ba
\lefteqn{
Z_W^{(n)} = 
{(-ig)^n \over N_c} \Tr (T^{a_1} \ldots T^{a_n})
\displaystyle\int d\tau_1 {p_1^{\mu_1} \over m} \ldots 
\displaystyle\int d\tau_n {p_1^{\mu_n} \over m} \times
} \nonumber \\
& & \times \theta (\tau_n - \tau_{n-1}) \ldots \theta (\tau_2 - \tau_1)
\langle A_{\mu_1}^{a_1} ({p_1 \over m} \tau_1) \ldots
A_{\mu_n}^{a_n} ({p_1 \over m} \tau_n) \rangle ~.
\ea
In the Euclidean theory we have, analogously:
\be
Z_{W E} \equiv {1 \over N_c} \langle \Tr [ W_{1E} (0) ] \rangle_E ~,
\ee
and
\ba
\lefteqn{
Z_{W E}^{(n)} = 
{(-ig)^n \over N_c} \Tr (T^{a_1} \ldots T^{a_n})
\displaystyle\int d\tau_1 p_{1E}^{\mu_1} \ldots 
\displaystyle\int d\tau_n p_{1E}^{\mu_n} \times
} \nonumber \\
& & \times \theta (\tau_n - \tau_{n-1}) \ldots \theta (\tau_2 - \tau_1)
\langle A_{(E) \mu_1}^{a_1} (p_{1E} \tau_1) \ldots
A_{(E) \mu_n}^{a_n} (p_{1E} \tau_n) \rangle_E ~.
\ea
Using Eq. (2.13), we can derive the following relation:
\ba
\lefteqn{
{\tilde{p}_1^{\mu_1} \over m} \ldots {\tilde{p}_1^{\mu_n} \over m}
\langle A_{\mu_1}^{a_1} ({\tilde{p}_1 \over m} \tau_1) \ldots
A_{\mu_n}^{a_n} ({\tilde{p}_1 \over m} \tau_n) \rangle =
} \nonumber \\
& & = {p_{1E}^{\mu_1} \over m} \ldots {p_{1E}^{\mu_n} \over m}
\langle A_{(E) \mu_1}^{a_1} ({p_{1E} \over m} \tau_1) \ldots
A_{(E) \mu_n}^{a_n} ({p_{1E} \over m} \tau_n) \rangle_E ~,
\ea
where, as usual, $p_{1E} = ({\bf p}_{1E}, p_{1E4})$ and
$\tilde{p}_1 = (\tilde{p}_1^0, {\tilde{\bf p}}_1) 
= (-i p_{1E4}, {\bf p}_{1E})$. If we define
\ba
Z_W \equiv h_M (p_1) ~~ &,& ~~ Z_W^{(n)} = h_M^{(n)} (p_1) ~,
\nonumber \\
Z_{W E} \equiv h_E (p_{1E}) ~~ &,& ~~ Z_{W E}^{(n)} = h_E^{(n)} (p_{1E}) ~,
\ea
from Eq. (2.36) we obtain
\be
h_E^{(n)} ({p_{1E} \over m}) = h_M^{(n)} (\tilde{p}_1) ~.
\ee
This relation is valid for every integer number $n$ and so we also have, 
more generally:
\be
h_E ({p_{1E} \over m}) = h_M (\tilde{p}_1) ~.
\ee
From the definitions (2.32) and (2.2), $h_M (p_1)$, considered as a 
function of a general four--vector $p_1$, is a scalar function constructed 
with the only four--vector $p_1$. In addition, $h_M (p_1)$ does not depend 
on the (positive) normalization of $p_1$: in other words, 
$h_M (\alpha p_1) = h_M (p_1)$ for every positive $\alpha$.
Therefore, $h_M (p_1)$ is forced to have the form
\be
h_M (p_1) = H_M (u_1^2) =H_M (1) ~,
\ee
where $u_1 = p_1 / \sqrt{p_1^2}$ ($u_1^2 = 1$).
In a perfectly analogous way, for the Euclidean case we have that:
\be
h_E (p_{1E}) = H_E (v_{1E}^2) =H_E (1) ~,
\ee
where $v_{1E} = p_{1E} / |p_{1E}|$ ($v_{1E}^2 = 1$).
Therefore, the first member of Eq. (2.39) is just equal to
$h_E (p_{1E}/m) = H_E (v_{1E}^2) =H_E (1)$
[observing that $(p_{1E}/m) / |(p_{1E}/m)| = p_{1E} / |p_{1E}|$],
and the second member is given by
$h_M (\tilde{p}_1) = H_M (\bar{u}_1^2) =H_M (1)$, where
$\bar{u}_1 = \tilde{p}_1 / \sqrt{\tilde{p}_1^2}$ ($\bar{u}_1^2 = 1$).
Then Eq. (2.39) implies that
\be
H_E (1) = H_M (1) ~.
\ee
That is, from Eqs. (2.37), (2.40) and (2.41):
\be
Z_{W E} = Z_W ~.
\ee
Combining this identity with Eq. (2.31), we find the following relation
between the amplitudes $g_M (\chi; ~t) = M (\chi; ~t)/Z_W^2$ and
$g_E (\theta; ~t) = E (\theta; ~t)/Z_{W E}^2$:
\ba
g_M (\chi; ~t)
\mathop{\longrightarrow}_{\chi \to i\theta}
g_M (i\theta; ~t) = g_E (\theta; ~t) ~;
\nonumber \\
{\rm or:}~~ g_E (\theta; ~t)
\mathop{\longrightarrow}_{\theta \to -i\chi}
g_E (-i\chi; ~t) = g_M (\chi; ~t) ~.
\ea
We have derived the relation (2.44) of analytic continuation for a 
non--Abelian gauge theory with gauge group $SU(N_c)$. It is clear, from the 
derivation given above, that the same result is valid also for an Abelian 
gauge theory (QED). We have thus completely generalized the results of Ref.
\cite{zfp97}, where the same relation (2.44) had been proven for an 
Abelian gauge theory (QED) in the so--called {\it quenched} approximation 
and for a non--Abelian gauge theory (QCD) up to the fourth order in the 
renormalized coupling constant in perturbation theory.
The approach adopted in Ref. \cite{zfp97} consisted in explicitly
evaluating the amplitudes $g_M (\chi; ~t)$ and $g_E (\theta; ~t)$, in the 
Minkowski and the Euclidean world, in some given approximation (such as the
{\it quenched} approximation) or up to some order in perturbation theory and 
in finally comparing the two expressions so obtained. Instead, in this 
paper we have given a general proof of Eq. (2.44), which essentially 
exploits the relation (2.13) between the gluonic Green functions in the two 
theories.

Therefore, it is possible to reconstruct the high--energy scattering 
amplitude by evaluating a correlation of two infinite Wilson lines forming a 
certain angle $\theta$ in Euclidean four--space, then by continuing this 
quantity in the angular variable, $\theta \to -i \chi$, where $\chi$ is 
the hyperbolic angle between the two Wilson lines in Minkowski 
space--time, and finally by performing the limit $\chi \to \infty$ (i.e.,
$\beta \to 1$). In fact, the high--energy scattering amplitude is given by
\ba
\lefteqn{
M_{fi} = \langle \psi_{i\alpha}(p'_1) \psi_{k\gamma}(p'_2) | M | 
\psi_{j\beta}(p_1) \psi_{l\delta}(p_2) \rangle } \nonumber \\
& & \mathop{\sim}_{s \to \infty}
-i \cdot 2s \cdot \delta_{\alpha\beta} \delta_{\gamma\delta}
\cdot g_M (\chi \to \infty; ~t) ~.
\ea
The quantity $g_M (\chi; ~t)$, defined by Eq. (2.1) in the Minkowski world, is 
linked to the corresponding quantity $g_E (\theta; ~t)$, defined by Eq. (2.7) 
in the Euclidean world, by the analytic continuation (2.44) in the 
angular variables.
The important thing to note here is that the quantity $g_E (\theta; ~t)$, 
defined in the Euclidean world, may be computed non perturbatively 
by well--known and well--established techniques, for example
by means of the formulation of the theory on the lattice. 
In all cases, once one has obtained the quantity $g_E (\theta; ~t)$, one 
still has to perform an analytic continuation in the angular variable
$\theta \to -i \chi$, and finally one has to extrapolate to the limit 
$\chi \to \infty$ (i.e., $\beta \to 1$). For deriving the dependence on $s$
one exploits the fact that both $\beta$ and $\psi$ (or equivalently $\chi$) 
are dependent on $s$. In fact, from $E = m/\sqrt{1 - \beta^2}$ and from
$s = 4E^2$, one immediately finds that
\be
\beta = \sqrt{ 1 - {4 m^2 \over s} } ~.
\ee
By inverting this equation and using the relation $\beta = \tanh \psi$,
we derive:
\be
s = 4 m^2 \cosh^2 \psi = 2 m^2 ( \cosh \chi + 1 ) ~.
\ee
Therefore, in the high--energy limit $s \to 
\infty$ (or $\beta \to 1$), the hyperbolic angle $\chi = 2\psi$ is 
essentially equal to the logarithm of $s$ (for a finite non--zero quark 
mass $m$):
\be
\chi = 2\psi \mathop{\sim}_{s \to \infty} \ln s ~.
\ee
As an example, we have shown in Ref. \cite{zfp97} how, using this 
approach, one can re--derive the well--known {\it Regge Pole Model}
\cite{Regge}. Of course, the most interesting results are 
expected from an {\it exact} non perturbative approach, for example by 
directly computing $g_E (\theta; ~t)$ on the lattice: a considerable 
progress could be achieved along this direction in the near future.

\vfill\eject

{\renewcommand{\Large}{\normalsize}
}

\vfill\eject


\begin{thebibliography}{99}
\bibitem{Nachtmann91}
O. Nachtmann, Ann. Phys. {\bf 209}, 436 (1991).
\bibitem{Meggiolaro96}
E. Meggiolaro, Phys. Rev. D {\bf 53}, 3835 (1996).
\bibitem{Verlinde}
H. Verlinde and E. Verlinde, Princeton University, report No. PUPT--1319
(revised 1993); hep--th/9302104.
\bibitem{zfp97}
E. Meggiolaro, Pisa preprint, IFUP--TH 10/96 (1996); hep--th/9602104;
Zeitschrift f\"ur Physik {\bf C} (1997), currently in press.
\bibitem{Cheng-Wu-book} 
H. Cheng and T.T. Wu, {\it Expanding Protons: Scattering at High 
Energies} (MIT Press, Cambridge, Massachussets, 1987).
\bibitem{Lipatov}
L.N. Lipatov, in {\it Review in Perturbative QCD}, edited by A.H. Mueller
(World Scientific, Singapore, 1989), and references therein.
\bibitem{Arefeva80}
I.Ya. Aref'eva, Phys. Lett. {\bf 93B}, 347 (1980).
\bibitem{Korchemsky}
G.P. Korchemsky, Phys. Lett. B {\bf 325}, 459 (1994);
I.A. Korchemskaya and G.P. Korchemsky, Nucl. Phys. {\bf B437}, 127 (1995).
\bibitem{Arefeva94}
I.Ya. Aref'eva, Phys. Lett. B {\bf 325}, 171 (1994);
{\bf 328}, 411 (1994).
\bibitem{Dosch}
H.G. Dosch, E. Ferreira and A. Kr{\"a}mer, Phys. Rev. D {\bf 50}, 1992 
(1994).
\bibitem{Regge}
T. Regge, Nuovo Cimento {\bf 14}, 951 (1959); {\bf 18}, 947 (1960).
\end{thebibliography}
\end{document}